\documentclass{PoS}

\usepackage{graphics}
\usepackage{rotating}

\newcommand\aproxgt{\mathrel{%
      \rlap{\raise 0.511ex \hbox{$>$}}{\lower 0.511ex \hbox{$\sim$}}}}
\newcommand\aproxlt{\mathrel{%
      \rlap{\raise 0.511ex \hbox{$<$}}{\lower 0.511ex \hbox{$\sim$}}}}

\newcommand\cyg{Cyg~X-1}

\title{Suzaku Observations of Cyg X-1}

\ShortTitle{Suzaku Observations of Cyg X-1}

\author{\speaker{Michael Nowak}\thanks{With the assistance of Paolo
    Coppi (Yale), John E. Davis (MIT Kavli Institute), Manfred Hanke
    (Universit\"at Erlangen-Nuremberg), Sera Markoff (University of
    Amsterdam), Katja Pottschmidt (CRESST/GSFC/UMBC), Sarah Trowbridge
    (MIT Kavli Institute), and J\"orn Wilms (Universit\"at
    Erlangen-Nuremberg).  Anything coherent and truthful herein is
    largely thanks to them, while any incoherent ramblings are wholly
    my own.  This work was supported in part by NASA Grants SV3-73016
    and NNX08AE23G.}\\ MIT Kavli Institute, Cambridge, USA\\ E-mail:
  \email{mnowak@space.mit.edu}}

\abstract{We present highlights from a series of four simultaneous
  Suzaku/Rossi X-ray Timing Explorer (RXTE) observations of the black
  hole candidate \cyg. We first briefly summarize several key results
  from our decade long RXTE monitoring campaign (which to date
  contains over 250 observations).  We then comment on challenges of
  analyzing the Suzaku data, i.e., improving the aspect correction
  beyond that of the existing tools, and quantitatively assessing
  pileup.  All of our Suzaku observations (one, by design) occurred at
  or very near orbital phase 0 (superior conjunction), and hence show
  evolution in color-color diagrams due to X-ray absorption by
  material from the wind of the secondary.  We present simple partial
  absorption models for this evolution. We then compare the Suzaku and
  RXTE data, and explicitly divide the Fe line region into narrow and
  broad components.  Both are required for the Suzaku data, and are
  seen to be consistent with the RXTE data.  These Suzaku observations
  occurred near historically hard, low flux states.  We present fits
  of the broad band spectra with a simple phenomenological broken
  powerlaw model, as well as a more physically motivated
  Comptonization model.  Whereas the former class of models described
  nearly all of the RXTE campaign better than any physical model, here
  the latter model is slightly more successful.  The Comptonization
  model, however, exhibits little evidence for a soft disk component,
  which formally corresponds to a small, inner disk radius.  Whether
  this is physical, due to unmodeled absorption, or is a calibration
  issue, remains an open question.}

\FullConference{VII Microquasar Workshop: Microquasars and Beyond\\
		 September 1 - 5, 2008\\
		 Foca, Izmir, Turkey}

\begin{document}

\section{Summary of RXTE Campaign}

\cyg\ is one of the best studied of the galactic black hole candidates
(BHC).  It has been a persistent X-ray source, exhibiting, very
broadly, spectrally hard to somewhat softer X-ray states.  (Whereas
the hard X-ray tail softens, it never completely vanishes for \cyg.)
In \cyg\ these spectral states occur over a factor of only a few in
X-ray flux.  A summary of the system properties can be found in
\cite{nowak:99a}, while a summary of the broad band spectral behavior
can be found in \cite{wilms:06a}, and references therein.  Whereas
\cyg\ exhibits an interesting variety of variability behavior (see
\cite{pottschmidt:00a,pottschmidt:02a,axelsson:05a,gleissner:04a,uttley:05a},
etc.) and radio behavior (see
\cite{gallo:03a,gleissner:04b,nowak:05a,wilms:06a}, etc.) correlated
with the X-ray spectra and flux, here we concentrate on the X-ray
spectral properties.

\begin{figure}
   \centerline{ \includegraphics[width=1.\textwidth]{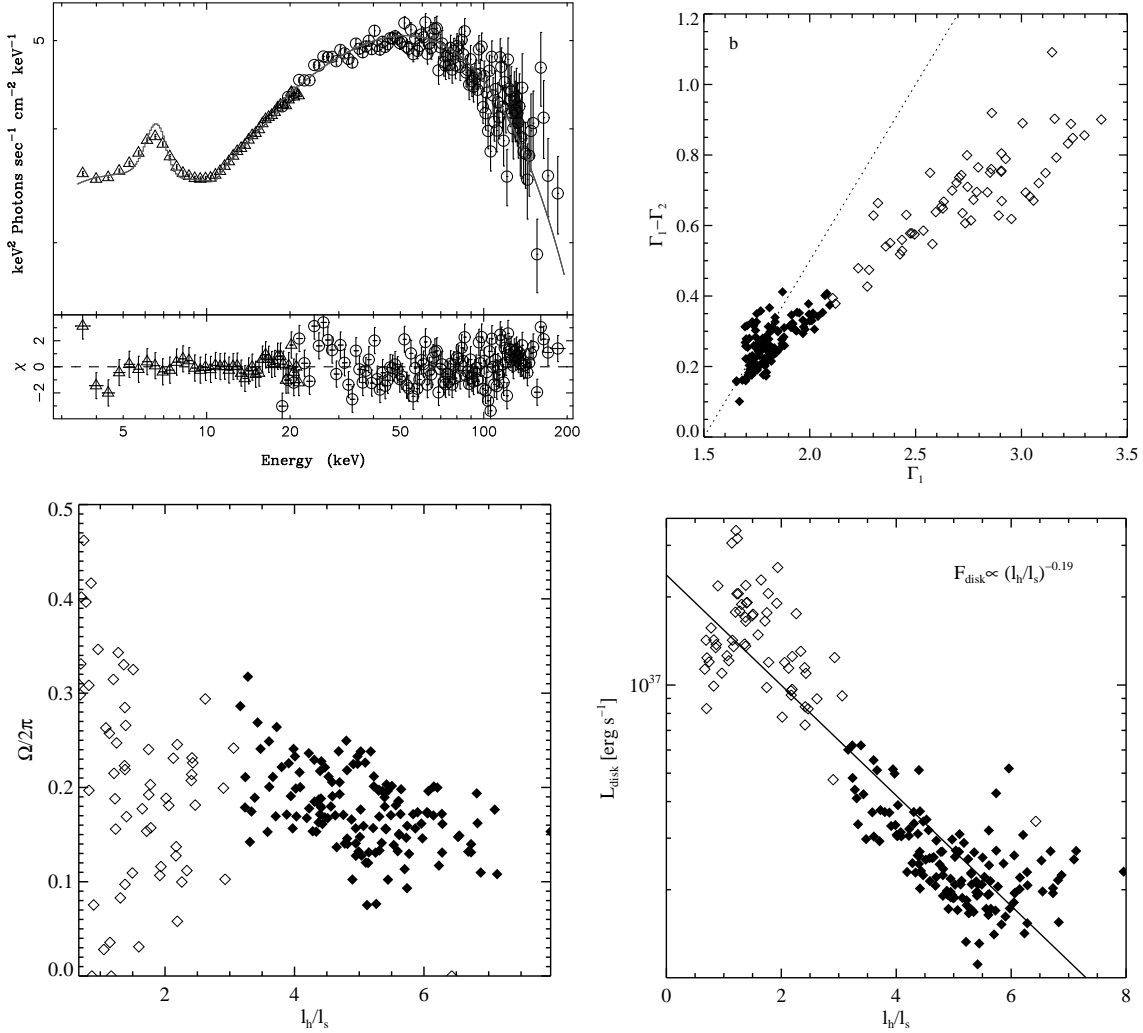}}
%\vskip -0.05 in
\caption{A brief summary of results from our monitoring campaign of
  \cyg\ (\protect{\cite{wilms:06a}}).  The upper left shows unfolded
  RXTE-PCA/HEXTE spectra (see also \protect{\cite{nowak:05a}}), fit
  with a simple absorbed, exponentially cutoff broken powerlaw plus
  broad gaussian line.  (This spectrum is among the softest seen in
  the campaign.) The amplitude of the powerlaw break is correlated
  with the soft powerlaw slope (softer corresponds to a greater
  break), as shown in the upper right.  Solid points are the
  \cyg\ hard state (as defined by \protect{\cite{remillard:06a}}), and
  hollow points are the \cyg\ soft state.  The lower left shows
  results from fitting the {\tt eqpair} Comptonization model of
  \protect{\cite{coppi:99a}}, which parameterizes spectral hardness
  with a ratio of coronal compactness to seed photon compactness
  ($\equiv \ell_h/\ell_s$).  Here the break between the soft and hard
  X-ray powerlaws are a combination of anti-correlations between the
  spectral hardness and reflection fraction (left) and amplitude of
  any additional, soft X-ray disk component (right).}
%\vspace*{-0.1 cm} 
\label{fig:summary}
\end{figure}

In Fig.~\ref{fig:summary}, we present some of the spectral highlights
from our RXTE monitoring campaign (\cite{wilms:06a}).  First, we see
that the broad band spectra are extremely well-described by a very
simple model: an absorbed, exponentially cutoff broken powerlaw, with
a broad gaussian line. The spectral break always occurs near 10\,keV,
with the degree of the break increasing for softer (typically
brighter) spectra.  More physically motivated Comptonization models
(\cite{ibragimov:05a,wilms:06a}) and X-ray jet models
(\cite{markoff:05a}) fit the data nearly as well.  The former fit the
soft X-ray power law with a combination of disk photons and
Comptonization, and the (exponentially cutoff) hard X-ray powerlaw
with Comptonization and reflection.  The jet models also utilize these
\emph{same} physical components, but also allow for jet synchrotron
radiation (usually in the soft X-ray), and jet synchrotron
self-Compton (usually in the hard X-ray).  For neither the
Comptonization nor jet models is there a single continuum component
underlying the Fe K$\alpha$ line region.

Prior studies have ascribed the $\Gamma_1$--$\Gamma_2$ correlation
predominantly to a hardness-reflection anti-correlation
(\cite{zdziarski:99a,ibragimov:05a}).  Although this effect is seen
somewhat by our campaign (\cite{wilms:06a}, Fig.~\ref{fig:summary}),
it is greatly reduced from prior claims (e.g., \cite{ibragimov:05a}).
A major difference between these studies is that we have allowed the
fitted seed photon temperature to be a free parameter, whereas other
studies have not (i.e., \cite{ibragimov:05a}).  Fig.~\ref{fig:summary}
shows that to some extent, a hardness-disk flux anti-correlation
subsumes a substantial fraction of any hardness-reflection
anti-correlation.  The ability of Suzaku to measure the soft X-ray
spectral regime is therefore seen to be crucial in further
constraining such models.

For all of our RXTE spectral fits, whether broken power law,
Comptonization, or jet, a broad gaussian is found near the expected
position of the 6.4\,keV, Fe K$\alpha$ line.  However, the width of
this line, and its correlations with spectral hardness, do depend upon
the exact fitted model.  (Broken power laws tend to produce the
narrowest lines; different Comptonization models, e.g., {\tt comptt}
vs. {\tt eqpair} yield differing line widths \cite{wilms:06a}.)
Furthermore, RXTE lacks the spectral resolution to decompose the
residuals in the line region into broad and narrow components.
Suzaku, on the other hand, does have sufficient resolution to
decompose the line into components, and again is seen to be crucial
for further constraining spectral models.

\section{Attitude Correction and Pileup Estimation for the Suzaku Observations}

\begin{figure}
   \centerline{
     \includegraphics[width=1.\textwidth]{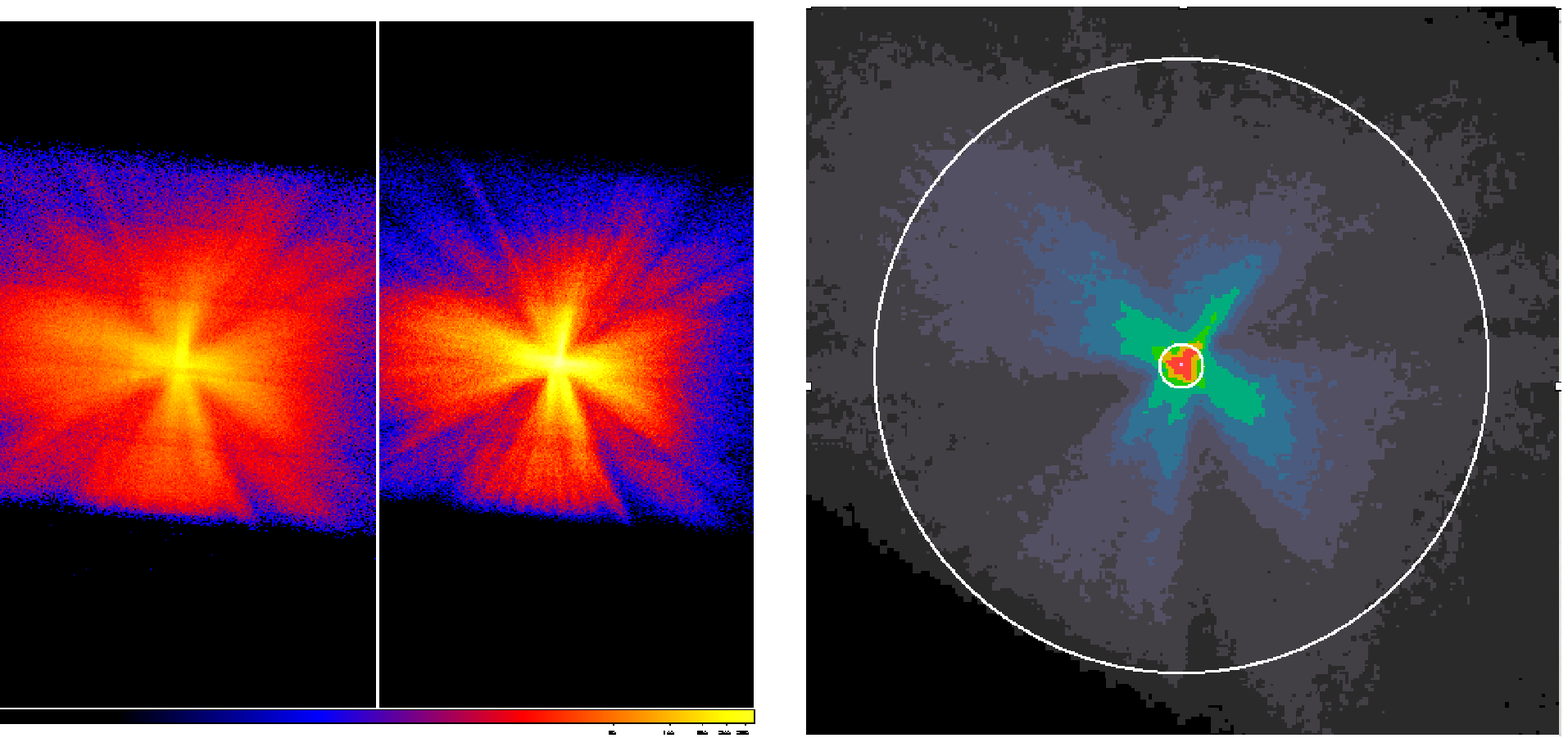}}
%\vskip -0.05 in
\caption{Left: An example of Suzaku attitude correction. The left half
  of the image shows a Suzaku PSF for \cyg\ using the standard
  attitude correction. The right half of the image shows the
  improvement with {\tt aeattcorr.sl}.  Right: Improving the attitude
  correction allows for a better estimate of pileup in the image.
  Here we show an image where discrete colors correspond to the
  minimum level of pileup in that region.  The outer white circle is
  the extraction region, while the inner white circle is the excluded
  region (with pileup fractions as high as 30\%). The average effective
  residual pileup level is less than 4\%.}
%\vspace*{-0.1 cm} 
\label{fig:pile}
\end{figure}

Before we can describe and fit the Suzaku spectra, we must first make
sure that we have minimized instrumental effects.  Most important
among these is the reduction of photon pileup, i.e., when more than
one photon lands in the same or adjacent pixels in one CCD readout
frame (see \cite{davis:01a} for a technical description).  These
``piled'' events are either lost (due to exceeding the event energy
threshold, or migrating to bad ``event grades''), or are read as a
higher energy photon, therefore distorting the spectrum.  (This
distortion can be especially problematic at high energies, as the
intrinsic photon count rate spectra tend to be $\propto E^{-1.7}$.)
In order to assess the degree of pileup, we must first have the most
accurate measure of the X-ray image.  This in turn requires an
accurate spacecraft attitude.  For Suzaku, the latter is affected by
``thermal wobbling'' of the spacecraft, induced during day/night
passages during the spacecraft orbit (\cite{uchiyama:08a}).

Whereas a statistical correction has been developed to correct the
spacecraft attitude (\cite{uchiyama:08a}), for bright sources (e.g.,
many AGN, and our \cyg\ data), the correction can be improved further
by measuring the mean position of the point spread function (PSF) on
short time scales, then refining the empirical correction of the
spacecraft attitude.  Such an algorithm, {\tt aeattcor.sl}, has been
developed by John Davis, and as shown in Fig.~\ref{fig:pile} it
further improves the \cyg\ image.

With this improved image, one can then calculate the mean counts per
$3\times3$ pixel region per CCD readout frame, and thereby arrive at a
quantitative assessment of the fractional pileup level in a given
image.  We have developed a tool ({\tt pileup\_estimate.sl}) to
automate this process.  It further allows one to create a region
filter that excludes the central piled regions, and returns an estimate
of the average pileup fraction in the remainder of the image.

These tools have been made available publicly\footnote{\tt
  http://space.mit.edu/CXC/software/suzaku/}.  For each of our
observations we exclude approximately 1/3 of the total counts from the
center of the (attitude corrected) image.  We estimate that any
residual pileup fractions are $\aproxlt 4\%$.

\section{The Suzaku Observations}

\begin{figure}
   \centerline{ \includegraphics[width=0.48\textwidth]{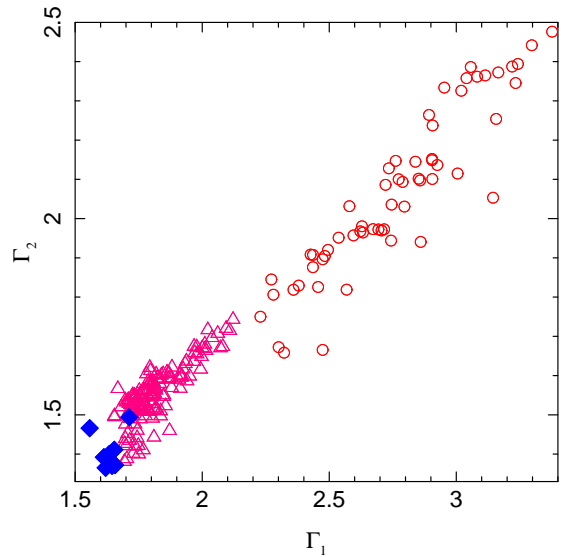}}
%\vskip -0.05 in
\caption{As discussed by \protect{\cite{wilms:06a}}, RXTE observations
  of Cyg X-1 are well-fit by an exponentially cutoff, broken
  powerlaw. The low energy power law ($\Gamma_1$) is strongly
  correlated with the high energy powerlaw ($\Gamma_2$).  The
  triangles correspond to the \cyg\ hard state, while the circles
  correspond to the \cyg\ soft state. The filled diamonds correspond
  to the RXTE observations simultaneous with our Suzaku observations.
  (Note that points are shown for eight individual color-color regions
  from the four separate observations.)}
%\vspace*{-0.1 cm} 
\label{fig:gamma}
\end{figure}

Over the course of a year and a half, we have obtained four
simultaneous Suzaku/RXTE observations.  (The most recent observation
in April 2008 was also performed simultaneously with Chandra,
XMM-Newton, INTEGRAL, and Swift.) In Fig.~\ref{fig:gamma} we show the
location in the $\Gamma_1$--$\Gamma_2$ diagram of broken powerlaw fits
to the simultaneous RXTE data.  We immediately see that these spectra
were among the spectrally hardest observations of the past decade.

\begin{figure}
   \centerline{ \includegraphics[width=0.48\textwidth]{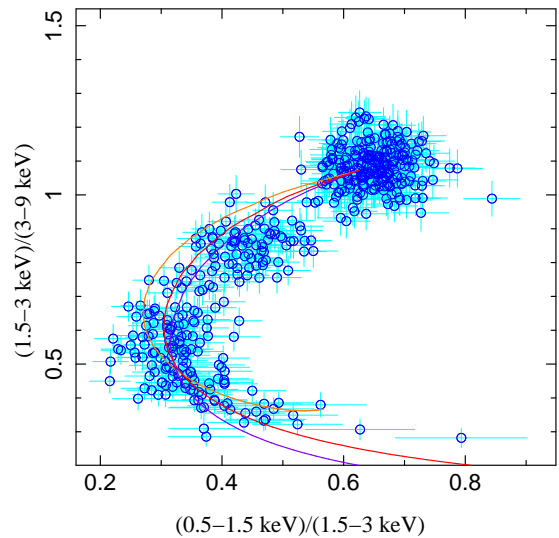}}
%\vskip -0.05 in
\caption{Color-color diagram for a Suzaku \cyg\ observation.  The
  evolution to the lower left corner of the diagram is consistent with
  increased absorption, while the rightward extending tail indicates
  \emph{partial} absorption.  The lines show partial covering models
  where only the absorption is changed.  From left to right,
  respectively, they represent: all channels having the same 82\%
  covering fraction, the two high energy channels having the same
  (100\%) covering fraction (the soft channel is $82\%$ covered), and
  the highest energy channel being completely uncovered (the soft
  channel is 81\% covered, the middle channel is 100\% covered). Three
  of our \cyg\ observations occurred near orbital phase 0, and show
  similar color-color diagrams.}
%\vspace*{-0.1 cm} 
\label{fig:color}
\end{figure}

Note that Fig.~\ref{fig:gamma} shows eight, rather than just four
points, as we have subdivided each observation by its location on a
color-color diagram defined by the Suzaku bandpasses (see
Fig.~\ref{fig:color}).  All of our observations show evolution towards
the lower left corner of these diagrams, with three of the four
observations rounding the corner and evolving to the right along the
bottom of the diagram.  These three observations (with durations of
0.2 in phase) overlapped with orbital phase 0 (i.e., superior
conjunction; the April 2008 observation was specifically scheduled for
that phase).  The fourth observation covered orbital phase 0.2--0.3.
Thus all observations were subject to a series of dipping events due
to absorption by the stellar wind of the secondary (see
\cite{balu:00a}).

We can model the color-color diagrams by fitting a spectrum to the
locus of points in the upper right hand of the plot, and then
multiplying this spectrum by a range of additional column densities
(above and beyond the Galactic column). This yields the
downward/leftward evolution on the diagram. In order to obtain the
evolution to the right along the bottom of the diagram, we need to
presume a partial covering fraction of 80\% (i.e., 20\% uncovered) for
at least the soft X-rays. Our best fit to the diagram, however, is for
the middle X-rays to be 100\% covered.  The uncovered soft X-rays,
rather than being wholly local to the source, is likely dominated by
the dust scattering halo (see \cite{predhel:95a}), a substantial
fraction of which is contained within the 2 arcminute Suzaku PSF.  We
note that Chandra observations of such dipping are more consistent
with a 95\% covering fraction (see \cite{hanke:08a}, and the
contribution by M. Hanke in this volume).

We have divided each observation into as many as three pieces - the
locus of points in the upper right, the downward slope to the left,
and the evolution to the right along the bottom of the diagram.  These
time intervals were defined by the Suzaku observations, and then RXTE
spectra were extracted from the intersections with those time
intervals.  Throughout the rest of this work, we only consider spectra
from the locus of points in the upper right.

\section{The Fe K$\alpha$ Line Region}

\begin{figure}
   \centerline{ \includegraphics[width=0.48\textwidth]{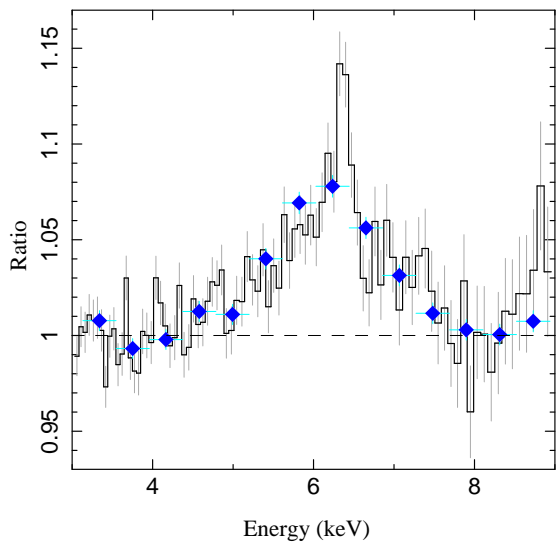}}
%\vskip -0.1 in
\caption{Fit residuals for RXTE (filled diamonds) and Suzaku
  (histogram), designed to emphasize features in the Fe~K$\alpha$
  region. A simple power law spectrum was fit to the 3.5--4.5 and
  7.5--8.5\,keV spectra, and then the full 3.5--9\,keV spectra were
  noticed without refitting the model.  The Suzaku data have been
  binned to half width half maximum (HWHM) of the spectral
  resolution.}
\vspace*{-0.1 cm} 
\label{fig:aline}
\end{figure}

We next turn to the line region of the data.  RXTE data have suggested
a broad Fe line (\cite{wilms:06a}), but questions have remained as to
contributions from narrow components, especially given the very modest
resolution of RXTE\footnote{Chandra resolves the narrow components of
  the line (see \cite{miller:02a}).  For Chandra, however, the broad
  components are difficult to assess with great accuracy (see
  \cite{hanke:08a}, and the contribution by M. Hanke in this
  volume).}. First, we wish to address the consistency between the two
detectors.  We do this by fitting a simple phenomenological model
(e.g., a powerlaw) simultaneously to the 3.5--4.5\,keV and
7.5--8.5\,keV data, and then noticing the ratio residuals in the full
3--9\,keV range.  This procedure does not necessarily produce an
accurate profile for any broad line in the 4--8\,keV region (it
produces perhaps the ``most optimistic'' estimate of such a line), but
it can show the consistency of the Suzaku and RXTE residuals.  We show
such ratio residuals in Fig.~\ref{fig:aline}.

There is indeed very good agreement between the Suzaku and RXTE
residuals, and both show broad deviations from a simple continuum,
consistent with a relativistically broadened line. (We have not,
however, included all the potential continuum complexity discussed
above, i.e., disk, Comptonization, synchrotron, synchrotron
self-Compton, and reflection). The Suzaku data also strongly suggest
narrow components.  A narrow emission feature is seen at 6.4\,keV, and
a possible absorption feature is seen at 6.7\,keV.  Based upon prior
Chandra observations at orbital phase 0 (\cite{hanke:08a}), both
features are expected.  The emission and absorption features have
comparable equivalent widths (absolute value $\approx10$--30\,eV), and
hence do not form the main contribution to the broad residuals seen in
the RXTE data (equivalent width $\approx 100$-$150$\,eV, if fitting
just a single component).

\begin{figure}
   \centerline{ \includegraphics[width=0.98\textwidth]{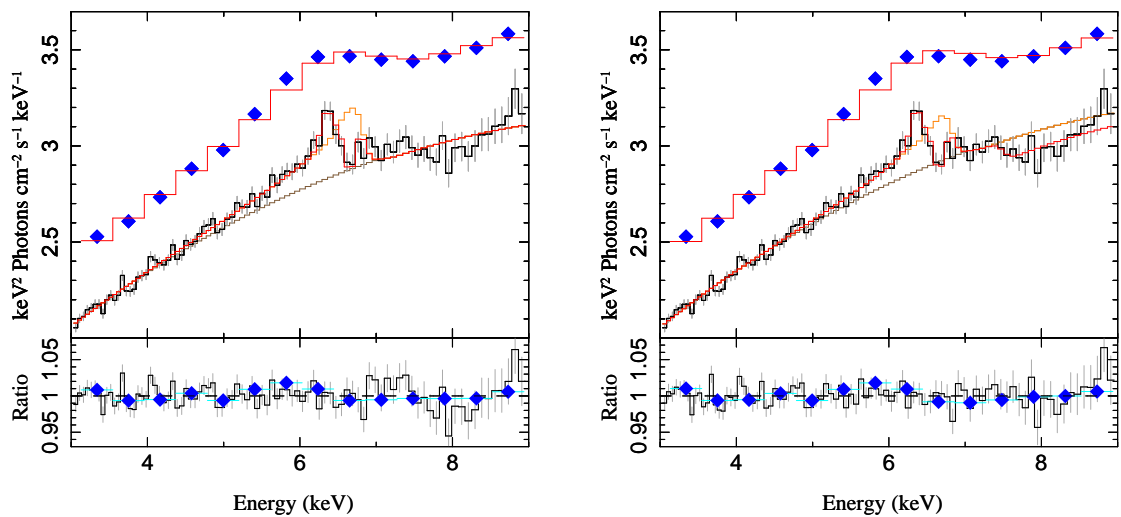}}
%\vskip -0.05 in
\caption{A simultaneous fit to the RXTE/Suzaku data in the 3.5--9\,keV
  range, consisting of an absorbed disk plus powerlaw, a
  relativistically broadened and a narrow Fe K$\alpha$ line, and a 6.7
  keV absorption line (left), plus an additional smeared edge (right).
  (Each model component is shown individually.) Such complexity is
  required for the Suzaku data, and is seen to give very good
  consistency with the RXTE data. Note that the data here are unfolded
  without reference to any model and only rely on the detector
  responses.}
\vspace*{-0.1 cm} 
\label{fig:line_fit}
\end{figure}

We see that \emph{at least} three components are required to fit the
broad line region residual in the RXTE data.  In
Fig.~\ref{fig:line_fit} we present such continuum plus multi-component
line fits.  Specifically, we include a relativistic {\tt diskline}
model, a narrow gaussian emission line, and a narrow gaussian
absorption line.  We show fits with and without an additional smeared
edge.  Both sets of fits require a broadened line component, with an
inner radius ranging from 10\,$GM/c^2$ to 60\,$GM/c^2$.  The
equivalent width of the broad line component is in the 40--130\,eV
range (and is 10\%--30\% lower if including a smeared edge).  It is
interesting to note that the peak in the blue wing of the line is
obscured by the absorption line at 6.7\,keV.  We have found, however,
that it is difficult to fit the width \emph{and} location of the RXTE
residuals without such a blue peak in the broad line component.

\section{Broad Band Fits}

\begin{figure}
   \centerline{ \includegraphics[width=0.98\textwidth]{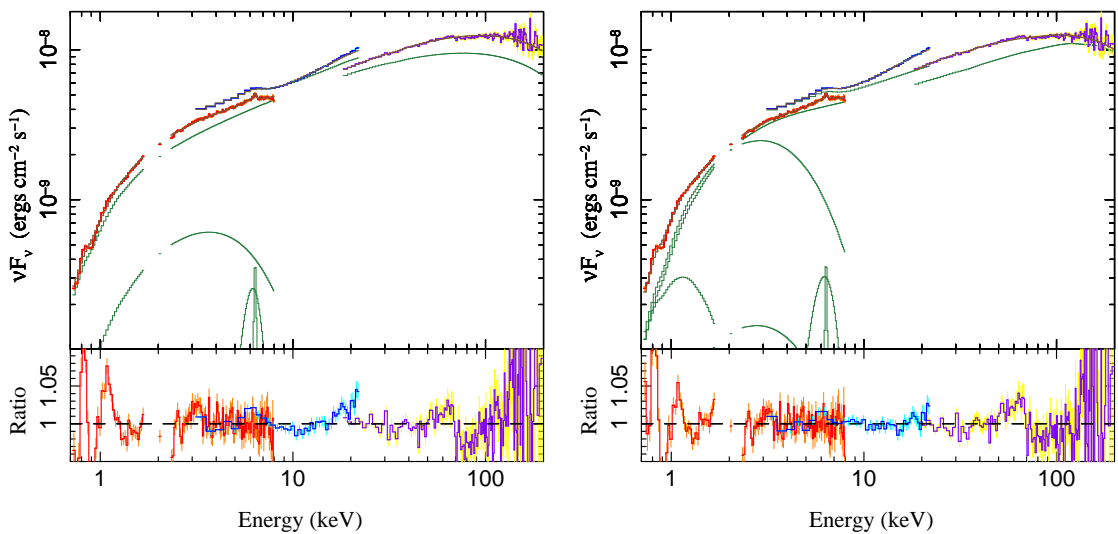}}
%\vskip -0.05 in
\caption{Broad-band \cyg\ spectra that have been unfolded solely using
  the detector response matrices (i.e., without regard to the fit
  model).  The Suzaku data from all available XIS have been summed and
  binned to HWHM of the detector, and we show the simultaneous
  RXTE-PCA and RXTE-HEXTE data.  (Model data have been ``unfolded''
  identically to the source data.) Left: absorbed, exponentially
  cutoff broken powerlaw, plus additional disk, and broad and narrow
  line components.  All (absorbed) model components are also shown
  individually, and we also show the broken powerlaw with and without
  the hardening above 10\,keV.  Right: The {\tt eqpair} Comptonization
  model fits, including additional unComptonized disk components (one
  tied to the seed photon temperature, and one free), broad and narrow
  lines, and reflection.  We also show the additional (absorbed) disk
  and line components individually, the (absorbed) seed photon
  spectrum (highest amplitude thermal bump in the figure), and the
  Compton spectrum without reflection.}
\vspace*{-0.1 cm} 
\label{fig:spectra}
\end{figure}

We now turn to the broad band spectra by considering models fit to the
0.7--9\,keV XIS data\footnote{For plotting purposes, all XIS spectra
  are summed, however, we fit each XIS data set individually.  The
  HXD-PIN data are in good agreement with the 20--70\,keV RXTE-HEXTE
  data, and do not alter the fits.  Consideration of the HXD-GSO
  background is a complex issue, and we defer discussion of these data
  to a forthcoming work.}, the 3--22\,keV RXTE-PCA data, and the
18--200\,keV RXTE-HEXTE data.  As for our RXTE campaign, we first
consider simple phenomenological fits consisting of an absorbed,
exponentially cutoff broken powerlaw.  Instead of adding a single
broad gaussian, we also include a narrow gaussian.  Additionally, we
include a soft disk component.  An example fit is shown in
Fig~\ref{fig:spectra}. 

Although this fit is relatively successful, several points are worth
noting.  For this, and all of our fits, there are sharp, $\approx 5\%$
residuals in the 0.7--1.2\,keV region, i.e., near absorption edge
features.  We fit absorption with {\tt tbnew}, a modification of the
model of \cite{wilms:00a}, which includes detailed structure at these
edges.  Altering absorption abundances, however, did not remove these
residuals.  We are unsure currently whether these residuals are
instrumental (e.g., contaminant unmodeled in the Suzaku spectral
response) or physical features (e.g., unmodeled line absorption or
emission).  The additional disk component is both of high temperature
and low amplitude.  Formally, low amplitude corresponds to a small
inner disk radius\footnote{Discussions of the inner disk radius
  \emph{not} receding as a BHC enters the low/hard state often have
  shown ratio residuals of the model-- excluding the disk component--
  in order to illustrate their points.  It is the relative
  \emph{weakness} of this ratio, however, that formally indicates a
  small inner disk radius, i.e., a non-recessed disk.}  Given the high
fitted disk temperature, here the disk amplitude implies an inner disk
radius of only $\approx 0.1\,GM/c^2$. A different (more heavily piled
up, even with removal of data from the center of the image) Suzaku
observation of \cyg\ recently has been described with a more complex
Comptonization model consisting of \emph{different} low \emph{and}
high energy Comptonized components (\cite{makishima:08a}).  Perhaps
the low normalization/high temperature disk component is indicating
the need for continuum models with such added complexity.

In Fig.~\ref{fig:spectra} we also show the same spectra fit with the
{\tt eqpair} Comptonization model, allowing for reflection, an
unComptonized disk component tied to the temperature of the seed
photons, and a second disk component with temperature left free (here,
$\approx 100$\,eV).  Overall, this fit is somewhat better than the
broken powerlaw model.  However, as for the broken powerlaw fit, any
soft excess (here dominated by the seed photons for Comptonization) is
both of high temperature, and low amplitude.  (To date, for these
historically hard ``low state'' observations, we have been unable to
find a satisfactory fit with low seed photon temperatures.)  Note also
that the fitted reflection fraction is $\approx 0.3$, i.e., counter to
any expectations from a hardness-reflection fraction
anti-correlation. We have simultaneous Chandra-HETG and XMM-RGS data
for one of our observations, thus, we are exploring whether the
inclusion of narrow absorption and emission components (e.g., from a
highly ionized wind) can fundamentally alter the broad-band properties
suggested by the above models.

%\bibliography{mnemonic,jw_abbrv,apj_abbrv,bhc,agn,diplom,inst,ns,conferences}
%\bibliographystyle{PoS}

\end{document}